\documentclass[twocolumn]{aastex63}

\usepackage{booktabs}
\usepackage{array}
\usepackage{multirow}
\usepackage{nameref}
\usepackage[T1]{fontenc}
\usepackage{amsmath}
\usepackage{hyperref}

\submitjournal{ApJ}

\newcommand{\review}[1]{{\color{black}{#1}}}
\shortauthors{To et al.}
\graphicspath{{./}{figures/}}

\begin{document}

\title{Systematic non-thermal velocity increase preceding soft X-ray flare onset: A large-scale Hinode/EIS study}

\author[0000-0003-0774-9084]{Andy S.H. To}
\affiliation{European Space Agency (ESA), European Space Research and Technology Centre (ESTEC), Keplerlaan 1, 2201 AZ Noordwijk, the Netherlands}

\author[0009-0000-5206-1030]{Abigail Burden}
\affiliation{University of St. Andrews, St. Andrews, UK}

\author[0000-0002-0665-2355]{Deborah Baker}
\affiliation{University College London, Mullard Space Science Laboratory, Holmbury St. Mary, Dorking, Surrey, RH5 6NT, UK}

\author[0000-0001-9597-3726]{Henrik Eklund}
\affiliation{European Space Agency (ESA), European Space Research and Technology Centre (ESTEC), Keplerlaan 1, 2201 AZ Noordwijk, the Netherlands}

\author[0000-0002-2189-9313]{David H. Brooks}
\affiliation{Computational Physics Inc., Springfield, VA, 22151, USA}
\affiliation{University College London, Mullard Space Science Laboratory, Holmbury St. Mary, Dorking, Surrey, RH5 6NT, UK}

\author[0000-0002-6835-2390]{Laura A. Hayes}
\affiliation{Astronomy \& Astrophysics Section, School of Cosmic Physics, Dublin Institute for Advanced Studies, DIAS Dunsink Observatory, Dublin D15XR2R, Ireland}

\author[0000-0002-0333-5717]{Juan Mart\'{i}nez-Sykora}
\affiliation{Lockheed Martin Solar \& Astrophysics Laboratory, 3251 Hanover Street, Palo Alto, CA 94304, USA}
\affiliation{SETI Institute, 339 Bernardo Ave, Mountain View, CA 94043, USA}
\affiliation{Rosseland Centre for Solar Physics, University of Oslo, P.O. Box 1029 Blindern, N-0315 Oslo, Norway}
\affiliation{Institute of Theoretical Astrophysics, University of Oslo, P.O. Box 1029 Blindern, N-0315 Oslo, Norway}

\author[0000-0002-0405-0668]{Paola Testa}
\affiliation{Harvard-Smithsonian Center for Astrophysics, 60 Garden Street, Cambridge, MA 02193, USA}

\author[0000-0003-4739-1152]{Jeffrey Reep}
\affiliation{Institute for Astronomy, University of Hawai’i, Pukalani, HI 96768, USA}

\author[0000-0003-0774-9084]{Miho Janvier}
\affiliation{European Space Agency (ESA), European Space Research and Technology Centre (ESTEC), Keplerlaan 1, 2201 AZ Noordwijk, the Netherlands}
\affiliation{Université Paris-Saclay, CNRS, Institut d'Astrophysique Spatiale, 91405, Orsay, France}

\author[0000-0001-7891-3916]{Shinsuke Imada}
\affiliation{Department of Earth and Planetary Science, University of Tokyo, 7-3-1, Hongo, Bunkyo-ku, Tokyo, Japan}

\author[0000-0002-4472-4559]{Julio Hernandez Camero}
\affiliation{University College London, Mullard Space Science Laboratory, Holmbury St. Mary, Dorking, Surrey, RH5 6NT, UK}

\author[0000-0003-3137-0277]{David M. Long}
\affiliation{Centre for Astrophysics \& Relativity, School of Physical Sciences, Dublin City University, Glasnevin Campus, Dublin, D09 V209, Ireland}
\affiliation{Astronomy \& Astrophysics Section, School of Cosmic Physics, Dublin Institute for Advanced Studies, DIAS Dunsink Observatory, Dublin D15XR2R, Ireland}

\author[0000-0001-8055-0472]{Teodora Mihailescu}
\affiliation{NASA Goddard Space Flight Center, Code 671, Greenbelt, MD 20771, USA}
\affiliation{Universities Space Research Association, Washington, DC 20024, USA}

\author[0000-0002-4433-4841]{Micah J. Weberg}
\affiliation{Space Science Division, Naval Research Laboratory, Code 7684, Washington, DC 20375, USA}




\begin{abstract}
Non-thermal velocities, derived from spectral line broadening, can provide crucial insights into plasma dynamics before and during solar flares. To systematically study the pre-flare phase, we constructed a Hinode/Extreme-ultraviolet Imaging Spectrometer (EIS) flare catalog of 1,449 flares from 2011-2024. This enabled a large-scale analysis of flare loop footpoint non-thermal velocity evolution across different flare magnitudes (C, M, X-classes). Analyzing \ion{Fe}{8}--\ion{Fe}{24} emission lines formed at log$(T/K) \sim5.7-7.3$ with piecewise linear fits in the pre-flare period, we find that non-thermal velocities consistently increase 4--25 minutes before GOES soft X-ray start in C and M-class flares. Onset timing patterns vary with flare magnitude: smaller flares show temperature-dependent progression, while larger flares exhibit more compressed, near-simultaneous onsets across temperatures. While our limited X-class sample (N=18) also show onset before GOES, larger statistics are needed to confirm its behavior. M-class flares show a systematic precursor non-thermal velocity onset $\sim$30--60 minutes before GOES peak. \review{In a subset of M-class flares (2011–2018), CME-associated events show earlier and more uniform precursor onsets (45–74 minutes before peak) than non-CME events, of which only some lines display a precursor, suggesting a strong link between extended pre-flare non-thermal broadening and successful eruptions}. This large scale study establishes pre-flare non-thermal velocity increase at footpoints as a common precursor observable before any X-ray signature.

\end{abstract}



\section{Introduction}\label{sect:intro}

Solar flares are explosive releases of magnetic energy in the solar corona that heat plasma to tens of millions of Kelvin and generating complex plasma dynamics including waves, turbulence, bulk flows, and a wide range of other phenomena. Non-thermal velocity, $v_{nt}$, derived from excess line broadening beyond thermal and instrumental effects, provides crucial information about unresolved plasma motions in flaring regions. These unresolved motions may originate from various physical mechanisms, such as turbulence generated during reconnection, wave propagation, pressure broadening, and unresolved bulk flows. The physical mechanisms driving non-thermal broadening in flares are highly case-dependent, varying significantly with atmospheric height, temperature, and local plasma conditions. Proposed explanations range from turbulent motion, superposition of plasma motions along the line of sight to Stark broadening and departures from ionization equilibrium~\citep{Milligan2011ApJ...740...70M,Polito2019ApJ...879L..17P}. These broadened spectral profiles are distributed throughout flaring structures, with the highest values typically observed in coronal loop tops during the late impulsive phase, and gradually decreasing along loop legs as flares decay~\citep{Warren2018ApJ...854..122W,French2020ApJ...900..192F,Stores2021ApJ...923...40S}.

Interestingly, $v_{nt}$ enhancements have been detected before the onset of both flare extreme-ultraviolet (EUV) and X-ray emissions. The timing of $v_{nt}$ enhancements varies significantly, appearing from minutes to tens of minutes~\citep{Harra2001ApJ...549L.245H, Harra2013ApJ...774..122H, Stores2021ApJ...923...40S, McKevitt2024ApJ...961L..29M}, or even an hour before X-ray flare signatures~\citep{Doschek1980ApJ...239..725D, Harra2009ApJ...691L..99H, Wallace2010SoPh..267..361W}, primarily near flaring footpoint regions. Other spectroscopic forms of pre-flare activity, such as bulk Doppler flows and intensity variability, have also been observed an hour prior to X-ray flares~\citep{Imada2014PASJ...66S..17I, Bamba2017ApJ...840..116B, Kniezewski2024ApJ...977L..29K}. Complementary to these coronal studies, observations from the Interface Region Imaging Spectrograph (IRIS; \citealt{DePontieu2014SoPh..289.2733D}) have revealed pre-flare chromospheric and transition region signatures. For instance, changes in e.g., \ion{Mg}{2}~h\&k spectra have been detected up to 40 minutes prior to flares, potentially linked to chromospheric heating events~\citep{Jeffrey2018SciA....4.2794J, Panos2020ApJ...891...17P, Woods2021ApJ...922..137W, Panos2023A&A...671A..73P,Zbinden2024A&A...689A..72Z}. \review{Large-scale statistical studies using UV and EUV imaging have further confirmed the systematic nature of pre-flare activity, with enhanced short-timescale variability in coronal emission serving as a strong discriminator between flare-imminent and flare-quiet active regions~\citep{Leka2023ApJ...942...84L}.} These studies suggest that the $v_{nt}$ increase could be observed for X- and M-class flares and occurs even before the reported X-ray onsets, which is the early appearance of very hot (>10 MK) plasma at or near flare footpoints occurring before the rise of hard X-ray signatures that signal collisional heating by non-thermal electrons~\citep{Hudson2021MNRAS.501.1273H,Battaglia2023A&A...679A.139B}. 

These pre-flare $v_{nt}$ enhancements suggest that a localized increase in non-thermal velocity near the flaring region may serve as an early indicator of impending flare activity. However, previous studies have focused on individual events or a handful of samples. This lack of systematic, large-sample analysis makes it difficult to determine whether observed pre-flare $v_{nt}$ behavior is representative or idiosyncratic, and limits our understanding of how non-thermal broadening characteristics might vary with flare energy or other properties. A comprehensive understanding of the spatial and temporal evolution of $v_{nt}$ across different flare magnitudes, enabled by a large sample size, is essential for identifying common mechanisms of energy release and transport during flares. 
In addition, statistical studies of stellar flares on solar-type stars have revealed remarkably similar power-law distributions in flare energy and duration relationships across a wide range of stellar types~\citep{Zhao2024ApJ...961..130Z}, suggesting common underlying physical processes. This convergence between solar and stellar flare properties makes large-scale solar studies particularly valuable, especially in light of upcoming missions such as MUSE (Multi-slit Solar Explorer; \citealt{DePontieu2020ApJ...888....3D}) and Solar-C EUVST (EUV high-throughput Spectroscopic Telescope; \citealt{Shimizu2020SPIE11444E..0NS}).

\begin{figure}

    \centering
    \includegraphics[width=\linewidth]{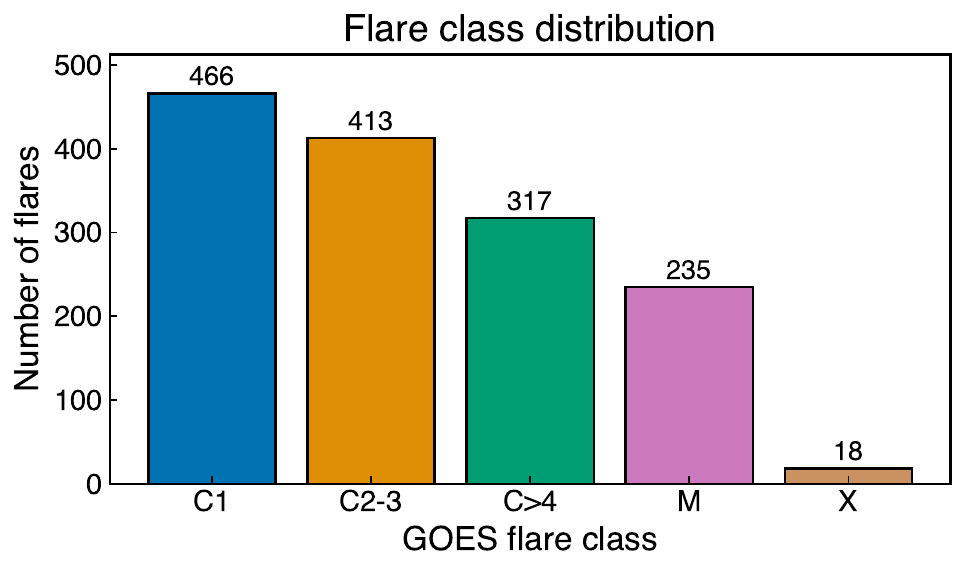}
    \caption{Flare class distribution analyzed in this study. We categorize the flares into 5 categories, C1, C$2-3$, C$>4$, M and X-class.}
    \label{fig:flare_stats}
\end{figure}

In this study, we address this gap by presenting a systematic analysis of flare footpoint $v_{nt}$ evolution by compiling and analyzing a large sample of 1,449 solar flares observed by Hinode/Extreme-ultraviolet Imaging Spectrometer (EIS; \citealt{Culhane2007Jun}) between 2011 Jan 1 and 2024 Nov 9. The start date was chosen to coincide with the availability of data from the Atmospheric Imaging Assembly (AIA; \citealt{Lemen2012Jan}) on board the Solar Dynamics Observatory (SDO; \citealt{Pesnell:2012}), enabling a systematic detection of flare ribbons. By analyzing emission lines covering a broad temperature range (\ion{He}{2}~256.32~\AA; $log(T/K)\sim4.7$), \ion{Fe}{8}~185.21~\AA\ to \ion{Fe}{24}~255.10~\AA; $log(T/K)\sim5.7-7.3$), we probe the non-thermal velocity evolution across different temperatures in the solar atmosphere with respect to flare intensity.

Leveraging these spatially and spectrally resolved observations enables us to answer key questions regarding flare-associated non-thermal velocities: (1) Do we observe a systematic increase in non-thermal velocity preceding the onset of flare-associated X-ray flux and energy injection? (2) How do non-thermal velocities evolve during different phases of solar flares? and, (3) Does this evolution pattern differ systematically across flare classes or eruption events?


\begin{figure*}
    \centering
    \includegraphics[width=\linewidth]{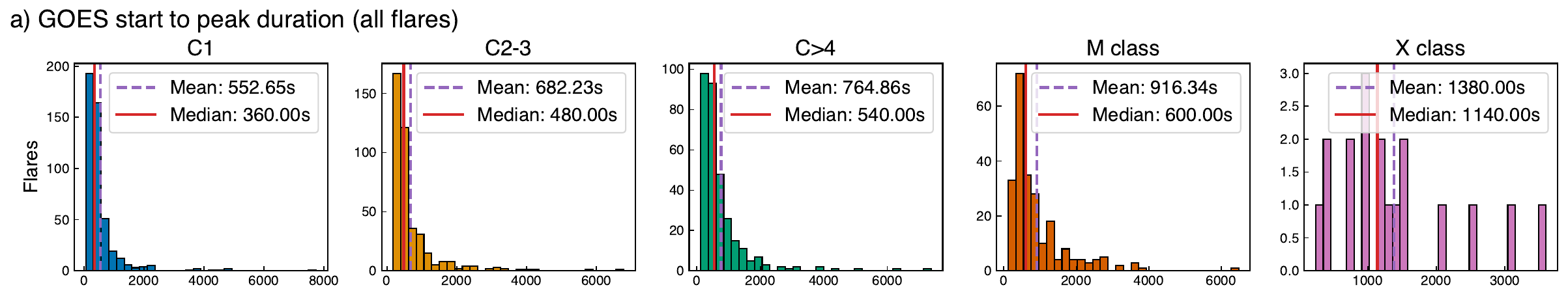}
    \includegraphics[width=\linewidth]{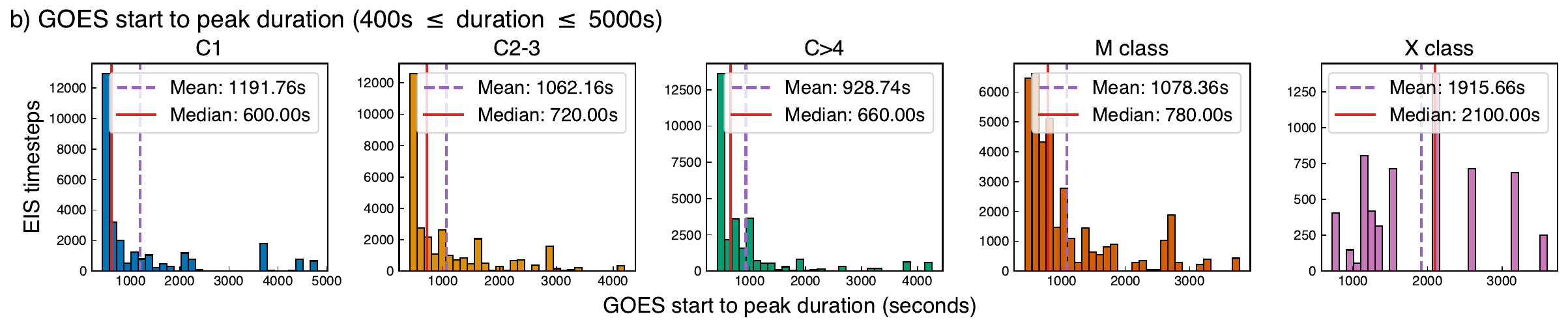}
    \caption{Statistics of the flare rise phase duration (GOES start to peak time difference) for the five flare categories. (a) Distribution of durations for all flares observed by EIS in the catalog. (b) Distribution of the number of individual EIS timesteps (y-axis) captured between 400~s and 5,000~s preceding the GOES flare peak, representing the data subset used for the pre-flare analysis in this study. The mean (dashed purple line) and median (solid red line) durations are indicated in each panel. The comparison shows that the pre-flare subset analyzed is slightly biased towards atypical rise durations compared to the overall sample.
}
    \label{fig:flare_start_stats}
\end{figure*}

\section{Observations and Data Analysis}\label{sect:obs}

To perform such a systematic study of pre-flare $v_{nt}$ evolution, we needed to compile a comprehensive EIS flare list spanning over a decade of observations. The difficulty arises from the need to cross-correlate multiple datasets, including identifying flare locations from the Geostationary Operational Environmental Satellite (GOES) flare list, and ensuring temporal and spatial overlap between the vast amount of EIS observations and flare events.

Through this systematic compilation process, we assembled a comprehensive dataset of 1,449 C-, M-, and X-class flares observed by Hinode/EIS between 2011 and 2024. We selected flares located between -800\arcsec\ and 800\arcsec\ solar-x to minimize projection effects associated with limb events. The distribution of flare classes in our sample is presented in Figure~\ref{fig:flare_stats}, consisting of 466 C1 flares (32\%), 413 C2-3  flares (29\%), 317 C$>4$ flares (22\%), 235 M-class flares (16\%), and 18 X-class flares (1\%). The following subsections detail the steps undertaken to create this dataset. 

\subsection{Determining flare location}
To establish a comprehensive flare list spanning 2011 to 2024, we first queried the Heliophysics Events Knowledgebase~\citep[HEK;][]{Hurlburt2012SoPh..275...67H} for the GOES flare start, peak, end time and magnitude. GOES start time is defined as the first minute in a sequence of 4 minutes of steep monotonic increase in the GOES 1-8~\AA\ X-ray flux\footnote{\url{https://www.swpc.noaa.gov/products/goes-x-ray-flux}, accessed 2025-06-10}. Since GOES does not provide accurate location of flares as it is an irradiance instrument, we implemented a standardized procedure to determine flare positions. Following \citet{Gallagher2002SoPh..209..171G} and the methodology employed by \texttt{SolarMonitor.org} to identify flare location, we generated difference images using 94~\AA\ images taken by SDO/AIA between the GOES flarelist-defined flare peak and start times to identify the flare loops. AIA data were prepared using the standard procedure available from \texttt{AIApy}~\citep{Barnes2020AIApy}. We applied Gaussian blur to these difference images, and the rough flare location was determined by the brightest pixel in the blurred difference image. Flares with no determinable or multiple locations were excluded from our analysis. This approach provides a reliable flare position. The derived locations are publicly available on \texttt{github.com/AbbyBurden/HEK-Flare-Positions}.

\subsection{GOES background removal}\label{sec:GOES_background_removal}
To ensure accurate flare classification and obtain a cleaner flare enhancement above the pre-flare corona, we applied additional processing to the GOES X-ray 2~s cadence data (XRS-B) beyond the standard catalog classes. For each flare, we reclassified them by implementing a rolling mean background subtraction method to remove the gradually varying non-flare component from the 1-8~\AA\ channel. Figure~\ref{fig:GOES_processing} in Appendix~\ref{appendix:GOES_bacground_removal} shows an example of the process. This approach provides a more robust measure of the actual flare enhancement above the pre-flare corona. The use of the science data and reclassification also accounts for the recent removal of a scaling factor by NOAA in the GOES data processing pipeline, which previously affected the reported flare magnitudes.

\subsection{EIS flare catalog}
We identified the observation time and field-of-view (FOV) of all Hinode/EIS raster observations between 2011 January 1 and 2024 November 9. These rasters were cross-referenced with the processed GOES flare list based on timing and the derived flare location. We required the flare location determined above to be within the EIS FOV $\pm20$\arcsec\ to account for pointing offset between AIA and EIS, and that the observation time overlapped with a window extending from 3 hours before the GOES flare start time to 3 hours after the GOES end time. In addition, we required that each EIS raster contain the \ion{Fe}{12}~195.12~\AA\ spectral window for the standard cross-correlation alignment between AIA and EIS, though our analysis encompasses emission lines from \ion{Fe}{8}~185.21~\AA\ to \ion{Fe}{24}~255.10~\AA\ to probe the $v_{nt}$ evolution across a wide temperature range ($log~T/K \sim 5.7-7.3$), as well as the \ion{He}{2}~256.32~\AA\ intensity to facilitate comparison with previous IRIS observations and to investigate signatures of chromospheric heating.

EIS coverage varies considerably across events, with some flares observed in multiple raster observations throughout their evolution while others have more limited temporal sampling (e.g., a single raster before the flare and none during or after). Despite this inherent characteristic, the large sample size accumulated over 14 years ensures sufficient statistics across different flare phases for our analysis. \review{This study focuses on the pre-flare phase's $v_{nt}$ evolution. To contextualize our findings, we first examine the typical GOES SXR rise phase duration within the broader EIS observations and check for bias in our pre-flare subset.} Figure~\ref{fig:flare_start_stats} shows the distribution of durations between the GOES-defined flare start and peak times. Panel (a) shows this for all flares observed by EIS in our catalog, while panel (b) focuses on the subset of EIS observations occurring between 400 and 5,000~s preceding the flare peak. Comparing the two panels, the median start-to-peak durations are systematically longer in row (b) across all flare classes (C1: 360$\rightarrow$600 s, C2--3: 480$\rightarrow$720 s, C>4: 540$\rightarrow$660 s, M: 600$\rightarrow$780 s, X: 1140$\rightarrow$2100 s), indicating a moderate bias toward \review{long rise-time events} in our pre-flare subset. This sampling bias likely arises from two factors. First, \review{long rise-time} flares present a larger temporal window for EIS observations, making them more likely to be captured by EIS during the pre-flare phase. Second, the EIS flare trigger mechanism, which has an average response time of $\sim2$~minutes~10~s after the SXR rise \citep{Brooks2023FrASS..1049831B}, may preferentially capture and sample \review{long rise-time events}. While our pre-flare analysis focuses on observations before the SXR rise, this bias means our dataset likely \review{undersamples flares with very short rise time} and may not be fully representative of the general solar and stellar flare population described in \citet{Reep2021SpWea..1902754R,Zhao2024ApJ...961..130Z}, where flare full-width-half-maximum durations show no correlation with GOES X-ray flux magnitude. However, flare duration does scale with total energy in both solar \citep{Toriumi2017ApJ...834...56T} and stellar flares \citep{Zhao2024ApJ...961..130Z}, making our findings particularly relevant to \review{longer rise time}, higher-energy events.

\begin{figure*}

    \centering
    \includegraphics[width=\linewidth]{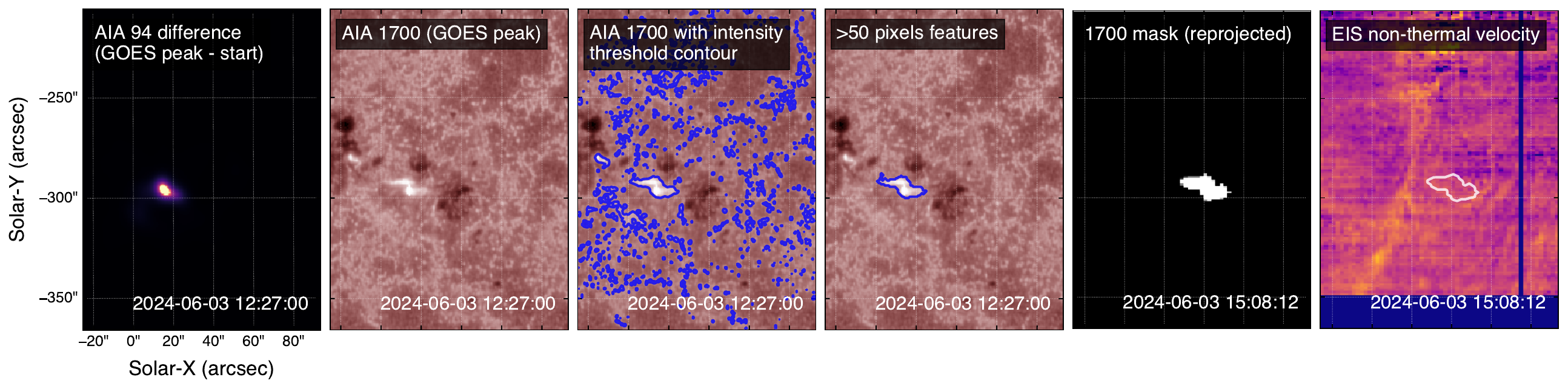}
    \caption{\review{
    Illustration of the steps taken to identify flare footpoints for EIS analysis detailed in Section~\ref{sect:obs}, using a M2.8 flare on 2024 June 3, with peak time at 12:27:00 UT. Left to right: AIA 94~\AA\ difference image between GOES peak and start time showing the flare location, AIA 1700~\AA\ flare footpoint image at GOES peak time, AIA 1700~\AA\ with intensity threshold contours (blue), filtered footpoint regions $>50$ AIA pixels, expanded by 3 AIA pixels and reprojected 1700~\AA\ mask to EIS WCS at 15:08:12 UT, and corresponding EIS \ion{Fe}{16}~262.98~\AA\ non-thermal velocity map with the footpoint mask overlaid (white contour).}
}
    \label{fig:method_illustration}
\end{figure*}

\begin{figure*}

    \centering
    \includegraphics[width=\linewidth]{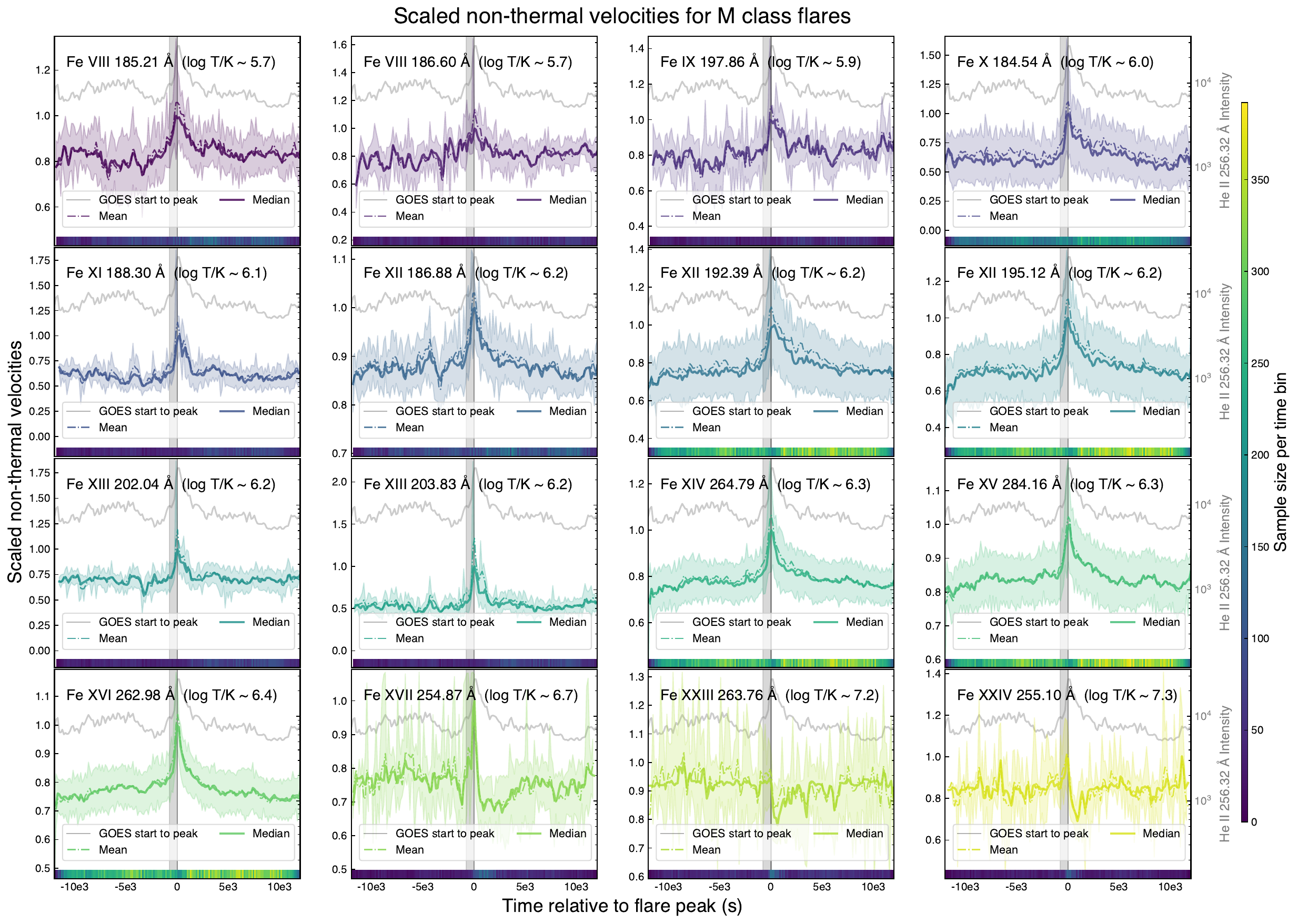}
    \caption{Non-thermal velocity evolution for M-class flares from 12,000~s before to 12,000~s (i.e. 200 minutes) after GOES X-ray peak. Solid varying line indicates the $v_{nt}$ median, dotted line indicates the mean, and the shaded area indicates the 25 and 75 percentile. The vertical shaded area indicates time between GOES X-ray start to peak for this flare category (Figure~\ref{fig:flare_start_stats}). The faint gray varying line above the $v_{nt}$ line indicates \ion{He}{2}~256.32~\AA\ intensity. The horizontal bar at the bottom of each subplot indicate how many vertical EIS columns or EIS timesteps went into the calculation of the respective temporal bin, with yellow indicating more time steps, purple indicating fewer. Emissions from \ion{Fe}{12}, XIV, XV and XVI are the better sampled ions across different flare classes. Data used to create this figure and similar plots for all flare classes are available at \href{https://doi.org/10.5281/zenodo.15613861}{doi:10.5281/zenodo.15613861}.}
    \label{fig:M_class}
\end{figure*}

To extract spectral information, we processed the EIS data using \texttt{EISPAC}~\citep{Weberg2023JOSS....8.4914W}. This included fitting the spectral lines within the Fe XII 195.12~\AA\ window and performing a cross-correlation co-alignment with AIA 193~\AA\ images corresponding to the EIS observation start time. To systematically pinpoint the locations where flare energy deposition is expected, we focus on the flare loop footpoints defined at the GOES SXR peak time. \review{Figure~\ref{fig:method_illustration} shows an example of the steps taken to identify flare footpoints and reproject to EIS observations.} Footpoint regions were isolated by generating masks from AIA 1700~\AA\ observations at GOES SXR peak time using a 99.5 percentile intensity threshold. We then labeled connected flare loop footpoints and retained only those larger than 50 AIA pixels to exclude small bright features. The resulting mask was expanded outward by 3 AIA pixels. Using SunPy \citep{Mumford2022zndo....591887M}, these AIA 1700~\AA\ masks were then reprojected to the EIS World Coordinate System (WCS) for each corresponding EIS observation time. The flare loop footpoint locations, defined at the GOES peak, therefore serve as a spatial reference throughout the analysis to capture any physical event that occurs at these locations in the pre-flare phase. By applying the same masks to EIS rasters before, during, and after the flare with differential rotation accounted, we ensure that spectral measurements consistently sample the same physical region across all flare phases.

To track the temporal evolution of $v_{nt}$ during flares, we used a column-by-column analysis approach to construct a flare $v_{nt}$ light curve. For each vertical column in an EIS raster (representing a single timestamp), we extracted all spectra that overlapped with the translated 1700~\AA\ footpoint mask. These spectra were then summed to generate a single spectrum for that time step of a single flare, and calibrated using the latest EIS calibration~\citep{DelZanna2025ApJS..276...42D}. We then fitted Gaussian profile(s) depending on the number of blends of each Fe line, and specifically two Gaussian profiles for \ion{He}{2}~256.32~\AA. Non-thermal velocity was calculated as the excessive width of the fitted lines in the summed spectrum after correcting for the instrumental width and thermal width using
\begin{equation}
    W_{total}^2 = 4\ln2\left(\frac{\lambda}{c}\right)^2(v^2_{t}+ v^2_{nt}) + W^2_{inst},
\end{equation}
where $W$ is the measured full width half-maximum (FWHM) of the spectral line, $\lambda$ is the wavelength of the peak, $c$ is the speed of light, $v_{t}$ is the thermal speed given by $\sqrt{{2k_BT_i/m_i}}$, where $k_B$ is the Boltzmann constant, $T_i$ is the ion temperature, and $m_i$ is the ion mass. $W_{inst}$ is the EIS instrumental FWHM stored in the EIS \texttt{hdf5} metadata taken from the \texttt{eis\_slit\_width} SSW routine \citep{Young2011EISNote7}. By taking the spectral summing approach, this significantly improves the EIS signal-to-noise while preserving temporal evolution information.

As quality control measures, we reject spectral fits where the uncertainty in the line intensity exceeded 10\% of the fitted value. After filtering, our final dataset consists of 19,658 EIS observations distributed across the 1,449 flares with 234,718 distinct timestamps, providing a robust statistical foundation for analyzing non-thermal velocity patterns across different flare magnitudes and evolutionary phases. While our catalog includes the hot Fe lines (\ion{Fe}{17}~254.87~\AA, \ion{Fe}{23}~263.76\AA, and \ion{Fe}{24}~255.10~\AA), these are typically noisier during the pre-flare phase and are predominantly sampled after the SXR peak, likely the effect of EIS flare trigger. Given the limited samples of X-class events in our dataset, we focus our analysis on C and M-class flares, though we present X-class results for completeness. The full EIS non-thermal velocity flare catalog is publicly available at \href{https://doi.org/10.5281/zenodo.15613861}{doi:10.5281/zenodo.15613861}.

\section{Results}

\subsection{Overview of non-thermal velocity evolution}

An example of the temporal evolution of the Fe $v_{nt}$ medians is shown in Figure~\ref{fig:M_class} (M-class), with the rest of the four flare classes (C1, C2-3, C>4, X) presented in Appendix~\ref{appendix:multipanel} (Figures \ref{fig:C1}--\ref{fig:X_class}). To create the temporal evolution $v_{nt}$ profiles, we performed a superposed epoch analysis where all flare $v_{nt}$ measurements were aligned to their respective GOES soft X-ray peak times (t = 0~s). We then binned the data into 150 s intervals, and computed the mean and median $v_{nt}$ values across all flares in each class at each time step, and subsequently normalized these ensemble statistics by their peak values. Vertical shades of gray in the figures indicate the median GOES start to peak time (start time indicated in Figure~\ref{fig:flare_start_stats}; row b). Background gray varying line indicates the \ion{He}{2}~256.32~\AA\ intensity. The \ion{He}{2}~256.32~\AA\ line, formed in the upper chromosphere/lower transition region serves as an indicator of heating events impacting the lower atmospheric layers, allowing us to compare roughly the timing of previous IRIS observations (e.g., \citealt{Panos2020ApJ...891...17P,Panos2023A&A...671A..73P, Zbinden2024A&A...689A..72Z}) with our coronal $v_{nt}$ measurements.

Non-thermal velocities across spectral lines and flare classes show a coherent trend -- it systematically increase prior to the GOES start time, always peak at the GOES flare peak at t=0~s, and then gradually decrease as flares decay. For example, as seen in Figure~\ref{fig:M_class} for M-class flares, the $v_{nt}$ in lines such as \ion{Fe}{10}~184.54~\AA, \ion{Fe}{12}~195.12~\AA, \ion{Fe}{15}~284.16~\AA\ and \ion{Fe}{16}~262.98~\AA\ (representing better sampled lines and a wide range of formation temperatures) all exhibit this pre-flare increase. This rise can sometimes begin as early as 3 hours before GOES peak, and the rate of increase often accelerate as the GOES median flare start time approaches. However, in this study, we focus on the $v_{nt}$ onset $\sim$1.5 hours before GOES peak.

\subsection{Quantitative analysis using piecewise linear fits}

To quantify the $v_{nt}$ increase before the GOES/SXR onset, we applied a systematic approach combining data smoothing and piecewise linear fitting. First, we smoothed the ensemble median $v_{nt}$ profiles using a centered rolling window to reduce noise while preserving the underlying trends. We then applied a piecewise linear fitting algorithm (\texttt{pwlf} package in \texttt{Python}) to these smoothed profiles over the time interval from 400~s to 5,000~s (i.e. $\sim$6--83 minutes) before the GOES peak (Figure~\ref{fig:M_class} and profiles for other classes in Appendix~\ref{appendix:multipanel}). 

\begin{figure}[h!]
    \centering
    \includegraphics[width=\linewidth]{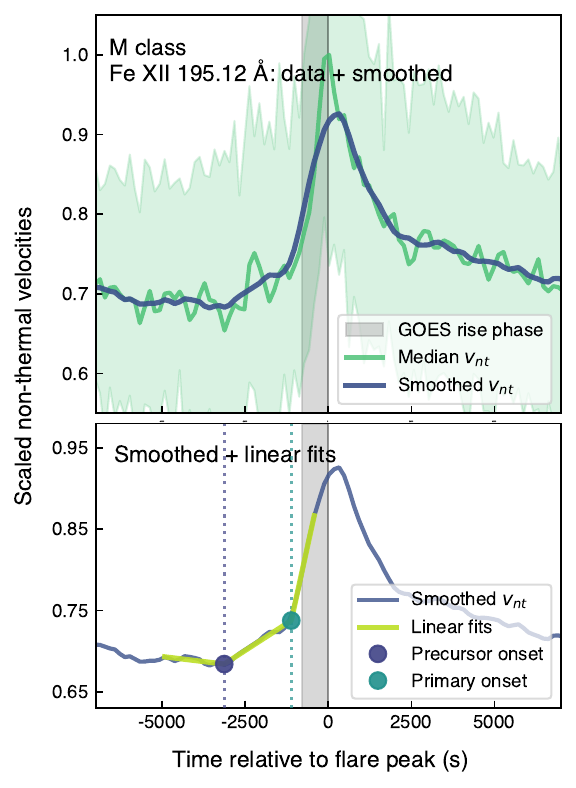}
    \caption{\review{Demonstration of the piecewise linear fit to determine non-thermal velocity onset times. Top panel shows the \ion{Fe}{12}~195.12~\AA\ $v_{nt}$ (green line), smoothed $v_{nt}$ using a smoothing window = 8 (purple line). Bottom panel shows the linear fits determined through the \texttt{PWLF} algorithm (light green line). First breakpoint (purple dot) indicates the precursor onset, second breakpoint indicates the primary onset (green dot).}}
    \label{fig:pwlf_demo}
\end{figure}

\review{Figure~\ref{fig:pwlf_demo} shows a demonstration of the piecewise linear fit algorithm to determine the onset times. The algorithm} determines the optimal placements of linear fits for a user-defined number of line segments by minimizing the least squares error. We specified 3 linear segments to capture the characteristic $v_{nt}$ evolution, and the algorithm automatically determines the optimal breakpoint locations that best fit the data. These breakpoints, where the slope changes between linear segments, define our onset times and provide measurements of when $v_{nt}$ begins to increase across different EIS EUV spectral lines~\citep{Jekel2018pwlf}. To assess the robustness of our onset time determinations, we performed a sensitivity analysis by varying the centered rolling window width from 5 to 10 time steps. The final onset time is determined as the mean of the breakpoint locations across all smoothing window sizes, with the standard deviation providing an estimate of the uncertainty. We disregard results with standard deviations >900~s.

\begin{figure*}

    \centering
    \includegraphics[width=\linewidth]{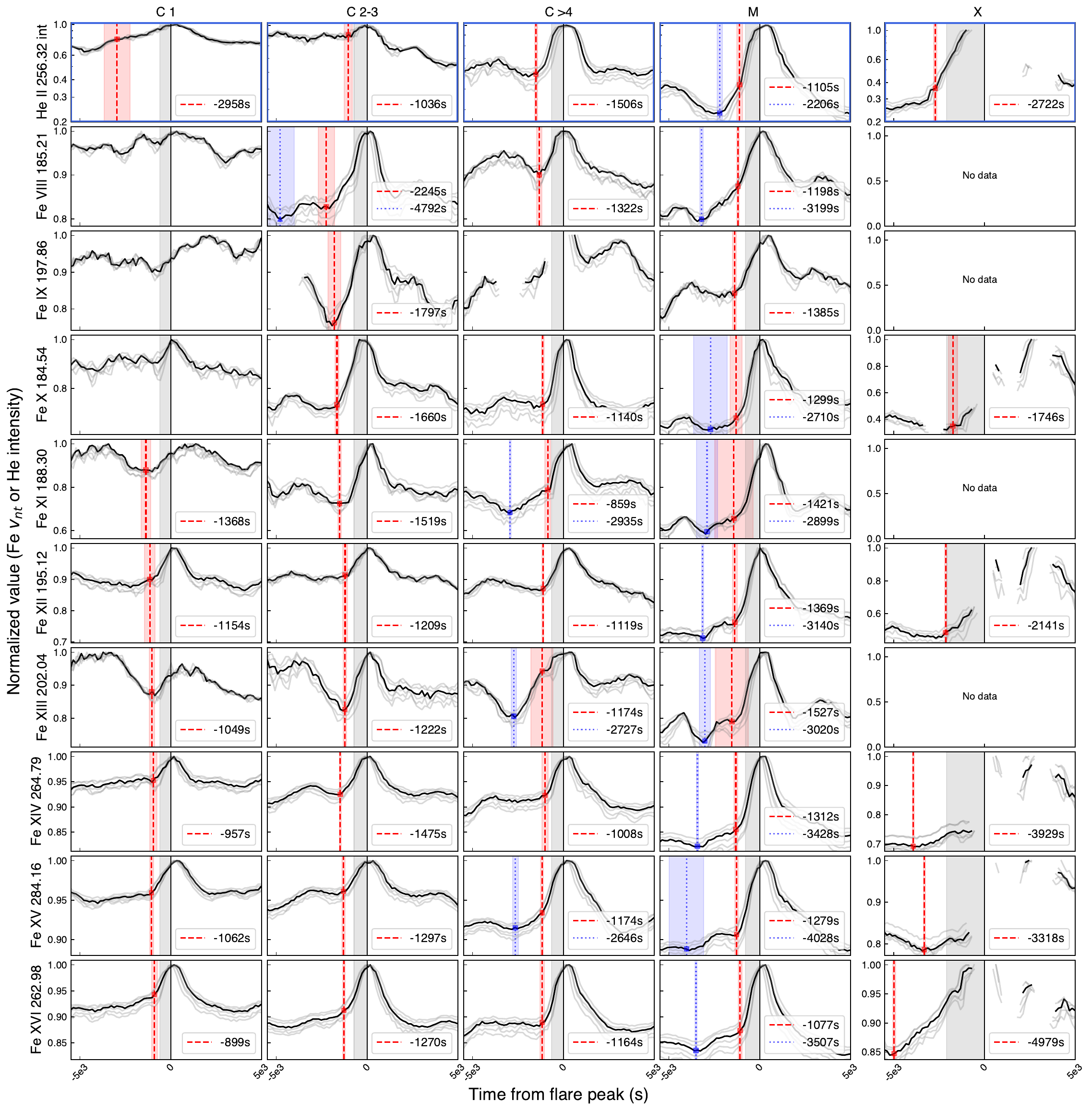}
    \caption{Normalized non-thermal velocities for different Fe emission lines and \ion{He}{2}~256.32~\AA\ intensity (top row) plotted against flare class and spectral line. Black vertical line at t=0~s indicates GOES flare peak time, shaded gray vertical areas indicates GOES flare start time to peak time. Red and blue dashed lines indicate the non-thermal velocity primary onset time with smoothing window = 8, and precursor onset time identified mostly in M-class, respectively. Grey varying lines above and below the black curves indicate the $v_{nt}$ evolution with different smoothing windows as part of our sensitivity analysis. The red and blue shaded areas around the dashed lines represent the standard deviation of the onset times.}
    \label{fig:onset_piecewise_linear_fit}
\end{figure*}

\begin{figure*}[ht!]

    \centering
    \includegraphics[width=\linewidth]{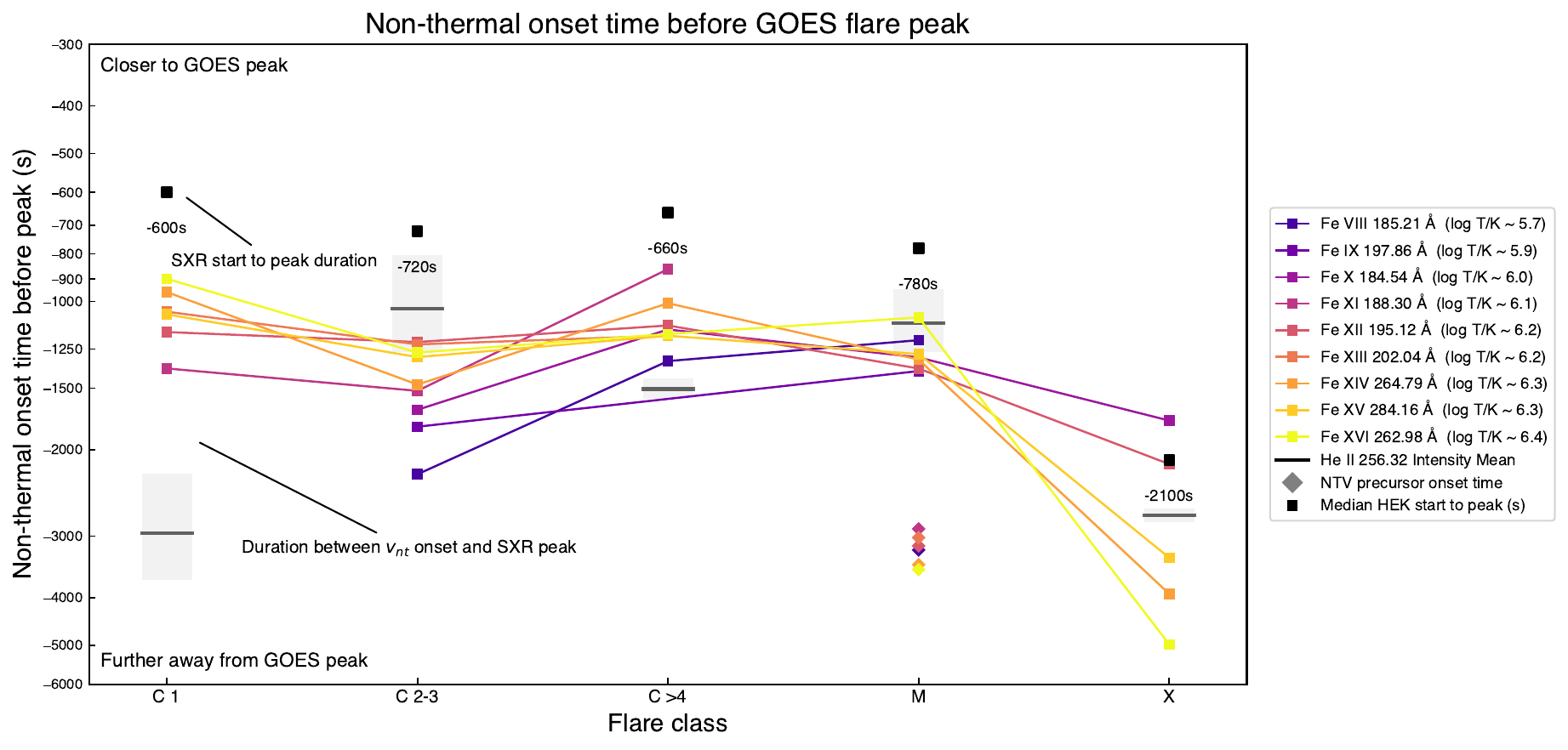}
    \caption{EUV non-thermal velocity onset time of the spectral lines in Figure~\ref{fig:onset_piecewise_linear_fit} (purple to yellow squares) compared to median GOES start time (black squares). Diamond shapes at M-class indicates precursor onset times illustrated as blue lines in Figure~\ref{fig:onset_piecewise_linear_fit}.
    GOES peak time is at t = 0~s. Gray horizontal lines and shades indicate the \ion{He}{2}~256.32~\AA\ intensity onset time. The evolution of onset times show a moderate dependence in C1 and C2--3 flares, but less so in C>4 and M-class flares.}
    \label{fig:NTV_onset_plot}
\end{figure*}

Figure~\ref{fig:onset_piecewise_linear_fit} presents the results of these fits using the normalized median profiles across five flare categories for a subset of emission lines from \ion{Fe}{8}--\ion{Fe}{16}. The black solid line indicates the median $v_{nt}$ smoothed using window width = 8, while the faint gray lines show the smoothed $v_{nt}$ using other window widths, illustrating the sensitivity analysis. The primary onset time (breakpoint closer to GOES peak at t = 0~s) identified by our analysis is marked with a red dashed line, while a secondary precursor onset time, observed most consistently in M-class flares, is indicated by a blue dashed-dotted line. The top row shows the \ion{He}{2}~256.32~\AA\ intensity, which signals possible chromospheric heating when intensity increases.

\subsection{Thermal dependency of primary $v_{nt}$ onset?}

\begin{figure*}[ht!]
    \centering
    \includegraphics[width=0.9\linewidth]{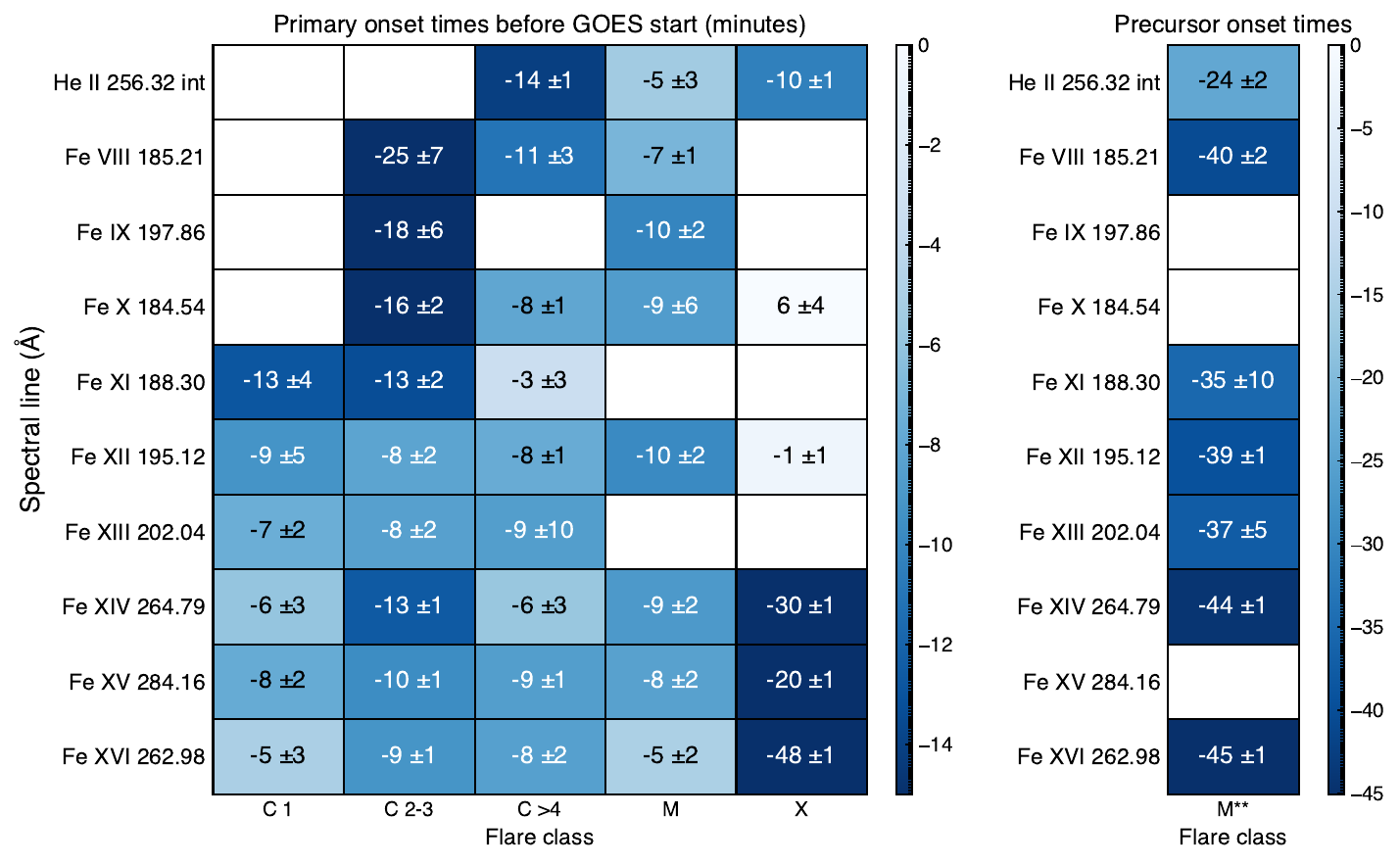}
    \caption{EUV non-thermal primary and precursor onset time (minutes) before GOES SXR start between flare categories and spectral line from \ion{Fe}{8} to \ion{Fe}{16}. Top row shows \ion{He}{2}~256.32~\AA\ intensity onset times. M** indicates the precursor onset time.}
    \label{fig:NTV_onset_table}
\end{figure*}

Examining the trends in $v_{nt}$ onset times across flare classes reveals several key patterns in the timing and temperature dependence of pre-flare non-thermal broadening. Through piecewise linear fits, as shown in Figure~\ref{fig:onset_piecewise_linear_fit}, and summarized in \ref{fig:NTV_onset_plot} (graphically) and \ref{fig:NTV_onset_table} (table and in minutes), we find that both Fe $v_{nt}$ enhancements and \ion{He}{2}~256.32~\AA\ intensity onset always precede the GOES SXR start time across all observed flare classes, with Fe $v_{nt}$ onset times ranging from 4 to 25 minutes before the GOES onset (C to M class flares).

In smaller flares (C1 and C2-3), we observe a tendency for a temperature-dependent progression. This is the clearest when comparing the coolest, though less frequently sampled, emission lines to the hotter lines. For example, in C2--3 flares, the \ion{Fe}{8}~185.21~\AA ($\sim$0.5 MK) line shows $v_{nt}$ enhancements beginning around 25 minutes before the SXR start, while the hotter \ion{Fe}{16}~262.98~\AA\ ($\sim$2.5 MK) line exhibits increased broadening only 9 minutes before. This pattern in smaller flares could suggest a more gradual and sequential energy release process, where reconnection energy deposition initially impacts the cooler, lower-lying coronal loops before propagating to regions of higher temperature emission.

Interestingly, hints of temperature dependence can also be observed in C>4 flares when we include the \ion{He}{2}~256.32~\AA\ intensity onset, which represents the coolest plasma in our analysis ($\sim$0.08 MK, upper chromosphere/lower transition region). In C>4 flares, \ion{He}{2}~256.32~\AA\ shows onset times around 14 minutes before the SXR start, slightly earlier than \ion{Fe}{8}~185.21~\AA\ (11 minutes), both precede other Fe coronal lines which cluster around 6-9 minutes before. However, this temperature progression becomes largely absent for the rest of the Fe lines in C>4 flares and for all lines in M-class flares. Here, the onset times at different temperatures are more compressed, appearing nearly simultaneous across the observed temperature range. For instance, in M-class flares, the onset times for lines from \ion{Fe}{8} to \ion{Fe}{16} are clustered within a narrow 5-minute window. With \ion{He}{2} showing similar timing to these Fe lines, this suggests that in more energetic events, the energy is deposited more explosively across a wider range of altitudes and temperatures, leading to a rapid, near-simultaneous response in the corona.

This interpretation is a bit more nuanced when we consider the observational sampling. The apparent temperature dependence in smaller flares is most pronounced in lines with lower number of EIS time steps. If we focus only on the emission of very well-sampled ions (\ion{Fe}{12}, \ion{Fe}{14}, \ion{Fe}{15}, and \ion{Fe}{16}) across all flare classes, the onset times are remarkably similar, with merely 5 minutes difference within them (except \ion{Fe}{14}~264.79~\AA\ in C2--3 class flares). This leads to a complementary picture for the primary non-thermal velocity onset times: the observed non-thermal velocity enhancement likely reflects the onset of vigorous, unresolved plasma motions associated with the initial energy release and transport processes. In this scenario, the primary onset marks the time when energy deposition has triggered large-scale plasma dynamics, whether through turbulent upflows from chromospheric evaporation, MHD wave propagation and dissipation, or other reconnection-driven processes.

For X-class flares, we find particularly early yet scattering onset times. Hotter emission lines such as \ion{Fe}{14}--\ion{Fe}{16} show onset times 20 to 48 minutes before the GOES start, while \ion{Fe}{10}~184.54~\AA\ and \ion{Fe}{12}~195.12~\AA\ show onset 6 minutes before to 1 minute after GOES start. Given our limited sample of X-class events (N=18), these timing results should be interpreted with caution compared to the more robust findings for C and M-class flares.

\begin{figure*}

    \centering
    \includegraphics[width=\linewidth]{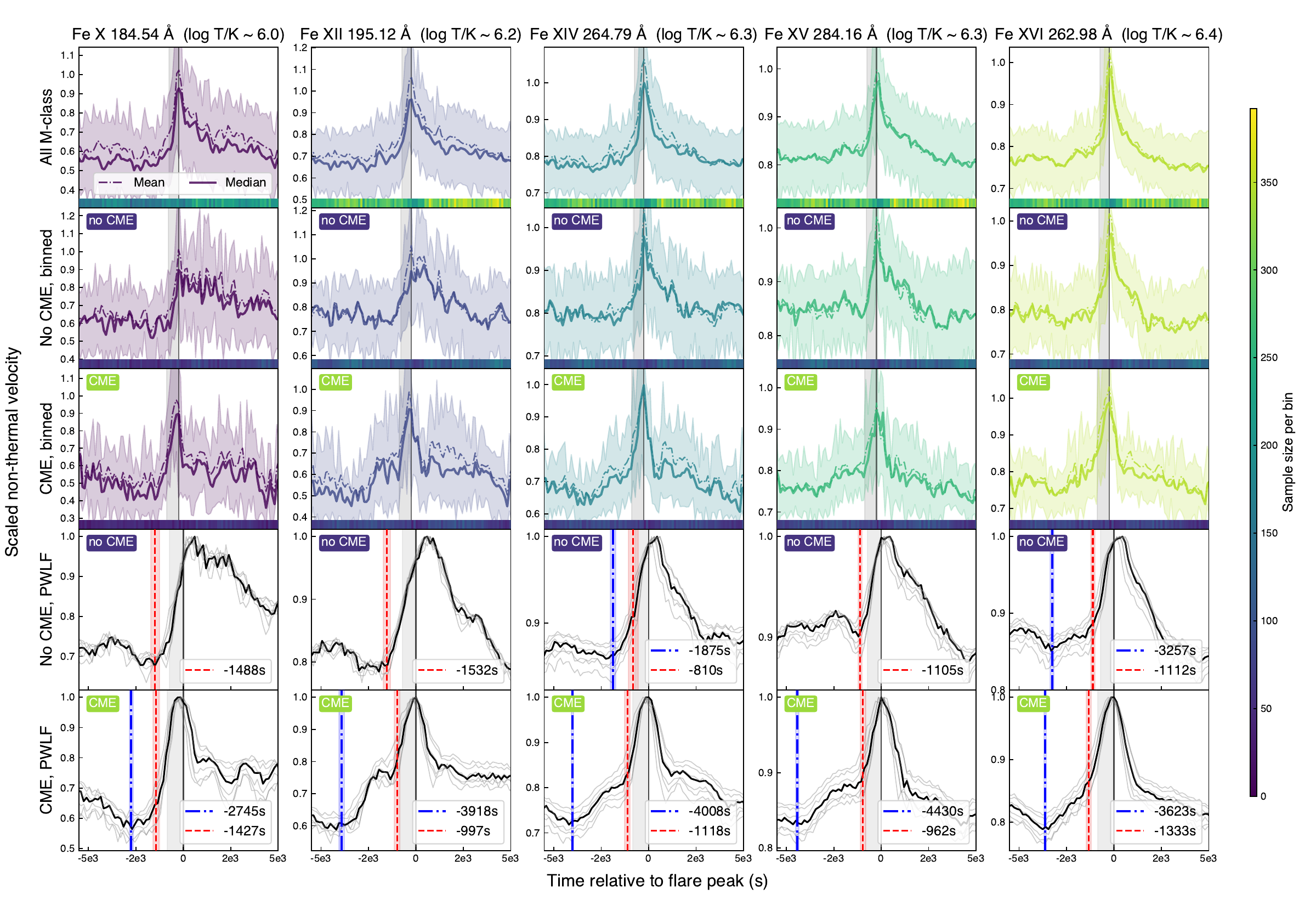}
    \caption{Non-thermal velocity analysis done on a subset of flares associated with CME from \citet{HernandezCamero2025ApJ...979...63H} between 2011 and 2018. M-class flares associated with CME shows much earlier precusor onset time (blue line).}
    \label{fig:cme_onset}
\end{figure*}

\subsection{Precursor onset}

Perhaps the most intriguing result is the systematic detection of a `precursor onset' specifically in M-class flares, where $v_{nt}$ enhancements first occur $\sim$24--45 minutes before the SXR start time (blue lines in Figure~\ref{fig:onset_piecewise_linear_fit}, diamond shaped scatter in Figure~\ref{fig:NTV_onset_plot}). Some precursor signatures can even be detected in C>4 flares. This suggests that previous findings from e.g.,  \citet{Harra2001ApJ...549L.245H,Harra2013ApJ...774..122H} is a common feature across M-class flares. Also, interestingly, the precursor onset shows a reversed temperature progression. Hotter emission lines e.g., \ion{Fe}{15}~284.16~\AA\ and \ion{Fe}{16}~262.98~\AA\ show $\sim$9--11 minutes earlier increase than \ion{Fe}{11}~188.30~\AA, $\sim$20 minutes earlier than the chromospheric/TR \ion{He}{2}~256.32~\AA\ intensity increase. The systematic nature of this early enhancement in M-class events, but more scarcely in C-class flares, as well as the reversed temperature dependency compared to the primary onset suggests that the physical processes contributing to the precursor onset are different from the main flare energy release, likely initiated higher up in the corona, with two possible interpretations: 
\begin{enumerate}
    \item First, it may represent a common initial energy release mechanism across all flare classes, but is simply below our detection threshold in smaller C-class events with current EIS sensitivity. This interpretation implies that with more sensitive instrumentation, we might be able to use these early $v_{nt}$ enhancements to predict eventual flare magnitude. 
    \item Alternatively, building on the work of ~\citet{Harra2013ApJ...774..122H, Woods2018ApJ...860..163W} who found pre-flare $v_{nt}$ enhancements associated with flux rope activation in specific events, this systematically observed precursor onset in M-class flares may be specifically linked to the processes like destabilization and gradual rise of a magnetic flux rope in the corona, resulting in the initiation of coronal mass ejections (for example a CME event from a pre-existing prominence destabilised by an emerging flux; e.g., \citealt{Janvier2023A&A...677A.130J}), which are more frequent with larger flares~\citep{Yashiro2005JGRA..11012S05Y}. \review{Recent high‑resolution Solar Orbiter observations suggest that such destabilization can proceed through a magnetic avalanche~\citep{Chitta2025arXiv250312235C}, where a cascade of small‑scale reconnection events in different parts of the system that progressively trigger one another, rapidly expanding from a few threads to the entire flux rope. In this scenario, our observed precursor $v_{nt}$ enhancements represent the spectroscopic signature of this avalanche‑like build‑up phase.}
\end{enumerate}

To test the second interpretation, we further divided our M-class flares into CME-productive and non-CME subsets using the automated CME–flare association catalog from \citet{HernandezCamero2025ApJ...979...63H}, which covers 2010 May 1 to 2018 December 31 and has a recall rate of $\sim$57\% for all frontside CMEs. \review{From this catalog, we identified 73 M-class flares and initially classified them as CME-productive or non-CME events. To improve the recall rate within this subsample, we visually inspected each event using SOHO/LASCO (Solar and Heliospheric Observatory satellite/Large Angle and Spectrometric Coronagraph Experiment; \citealt{Brueckner1995SoPh..162..357B}) and STEREO/SECCHI coronagraph data~\citep[Sun–Earth Connection Coronal
and Heliospheric Investigation;][]{Howard2008SSRv..136...67H}. This verification yielded two distinct M-class flare subsets, based on EIS observations within the 5,000 s window preceding the SXR peak: (i) 31 flares associated with CMEs (2,684 EIS timesteps) and (ii) 42 flares with no CME detected (3,741 EIS timesteps).}

Figure~\ref{fig:cme_onset} presents this comparative $v_{nt}$ analysis of $v_{nt}$ evolution patterns between the two populations, revealing distinct patterns between CME-productive and non-CME flare populations. We focus on the \ion{Fe}{10}~184.54~\AA, \ion{Fe}{12}~195.12~\AA, \ion{Fe}{14}~264.79~\AA, \ion{Fe}{15}~284.16~\AA, and \ion{Fe}{16}~262.98~\AA\ lines that are better sampled to show continuous $v_{nt}$ evolution in both CME and no-CME subsets. \review{Now, non-CME flares show inconsistent precursor onset behavior, with identifiable precursor onset limited to only the \ion{Fe}{14}~264.79~\AA\ and \ion{Fe}{16}~262.98~\AA\ lines. In contrast, CME-productive flares demonstrate consistently earlier precursor onset across multiple emission lines spanning a wide temperature range, occurring 45–74 minutes before the GOES peak compared to 31–54 minutes for the non-CME subset.}

\section{Discussion}

\subsection{Physical implications of pre-flare non-thermal velocities}

Our systematic analysis was enabled by the creation of a comprehensive EIS flare catalog spanning 2011-2024, containing 1,449 flares observed by Hinode/EIS. The results reveal a clear pattern: non-thermal velocity enhancements at flare loop footpoints consistently precede GOES soft X-ray signatures across all flare classes. These $v_{nt}$ increases are detectable in the \ion{Fe}{8}--\ion{Fe}{16} emission lines (formed at log(T/K)$\sim$5.7--6.4) up to 25 minutes before the onset of SXR, with onset timing patterns that vary systematically with flare magnitude. This finding establishes the primary $v_{nt}$ increase before SXR as a common feature in flares, suggesting a common initial stage of energy release whose characteristics depend on flare magnitude. The temporal sequence might point to processes that contribute to excessive coronal line broadening as fundamental to understanding the triggers for reconnection~\citep{Polito2019ApJ...879L..17P,McKevitt2024ApJ...961L..29M}.

The onset timing patterns we observe provide important clues about energy release mechanisms, revealing different behaviors across flare magnitudes. In smaller flares, the tendency for cooler lines to show earlier primary onset enhancement suggests a more sequential process. This could reflect spatial and temporal propagation of energy deposition and plasma heating, where cooler plasma is heated first and subsequently evaporates to fill coronal loops, creating the observed sequence from cooler to hotter spectral line signatures.  However, it is difficult to reconcile that the energy release would begin 25 minutes before the main phase in flares as small as C2--3. If this represented the true onset of energy release, one would expect M-class flares, with their more explosive nature, to show a rapid $v_{nt}$ increase 25 minutes before GOES onset across all Fe lines, yet we observe a narrow temporal spread between spectral lines for M-class flares. A more convincing interpretation emerges when we consider the near-simultaneously observed $v_{nt}$ enhancements in well-sampled lines across all flare classes. This suggests that the primary $v_{nt}$ onset reflects more rapid plasma dynamics, whether from chromospheric evaporation, wave dissipation, or other reconnection-driven processes that transported across our temperature range faster than our observational cadence, and the temperature dependency in lower class flares are due to sampling and analysis limits. Nonetheless, these findings are consistent with flare models where energy release impacts the lower atmosphere and can either propagate sequentially in smaller events or trigger more explosive, widespread responses in larger ones (e.g., \citealt{Jeffrey2018SciA....4.2794J,DiazBaso2021A&A...647A.188D}).

Our findings complement recent studies of `hot onset' flare phenomena~\citep{Hudson2021MNRAS.501.1273H, Battaglia2023A&A...679A.139B}, which show that very hot plasma (10--15 MK) appears at flare loop footpoints before significant hard X-ray emission. The systematic detection of enhanced $v_{nt}$ preceding both these hot thermal signatures and the SXR rise suggests a causal relationship: early turbulent motions causing the primary onsets directly contribute to plasma heating before the impulsive phase of a flare. This interpretation can be viewed alongside our \ion{He}{2}~256.32~\AA\ intensity measurements and other chromospheric/transition region studies. While \ion{He}{2}~256.32~\AA\ timing is not consistently earlier than coronal signatures across all flare classes (for example, in M-class flares \ion{He}{2} shows similar timing to Fe lines), previous IRIS observations from
\citet{Jeffrey2018SciA....4.2794J} demonstrated that turbulent velocity fluctuations in the lower solar atmosphere (at $\sim$80,000 K) precede GOES flare onset, showing non-thermal line broadening that peaks before the intensity rise and then oscillates with $\sim$10 s periods. Similarly, recent studies by \citet{Panos2020ApJ...891...17P} and \citet{Zbinden2024A&A...689A..72Z} have identified spectral signatures in the chromosphere and transition region tens of minutes before flare onset. When viewed together with our \ion{He}{2} analysis, these multi-instrument observations suggest that energy release processes impact multiple atmospheric layers, with the observable signatures depending on the energy deposition rate and flare magnitude rather than following a strict sequential progression from lower to higher temperatures.

\subsection{M-class precursor onset and CME initiation}

An important result of our study is the systematic detection of a `precursor onset' specifically in M-class flares, where $v_{nt}$ enhancements occur $\sim$25--45 minutes before the GOES start time (blue line in Figure~\ref{fig:onset_piecewise_linear_fit}). This early activity is less consistently observed in C-class flares. This suggests that M-class flares undergo a prolonged preparatory phase before their explosive main energy release. \review{Furthermore, our analysis of M-class flares associated with CME (Figure~\ref{fig:cme_onset}) reveals even earlier $v_{nt}$ and more consistent enhancements compared to confined M-class flares}, providing strong evidence linking early footpoint non-thermal broadening with successful eruptions.

Building on work by \citet{Harra2013ApJ...774..122H}, \citet{Woods2018ApJ...860..163W}, and \citet{McKevitt2024ApJ...961L..29M}, who linked pre-flare $v_{nt}$ to release of free magnetic energy and flux rope activation to the increase of $v_{nt}$, we propose that these early $v_{nt}$ trace the initial, gradual destabilization of the magnetic configuration leading to eruption. \review{\citet{Chitta2025arXiv250312235C} suggests that such destabilization can proceed through a magnetic avalanche, in which small‑scale reconnection events in one location trigger others, rapidly expanding from a few magnetic threads to the entire flux rope. Recent 3D MHD simulations of avalanche processes by e.g., \citet{Cozzo2023A&A...678A..40C, Cozzo2024A&A...689A.184C} have shown that small-scale magnetic reconnection during flux tube fragmentation produces distinctive spectroscopic signatures, including enhanced non-thermal broadening. These suggest that our observed precursor onsets trace the spectroscopic signature of this progressively more energetic destabilization. We therefore suggest a three‑stage process:}
\begin{enumerate}
    \item \textbf{Initial slow reconnection ($\sim$30--60+ minutes before M-class flare peak):} Driven by physical processes like emerging flux (which can induce small-scale heating, see \citealt{Yadav2023ApJ...958...54Y}), photospheric motions and/or flux cancellation, free magnetic energy accumulates and small-scale reconnection begins. For example, soft X-ray sigmoid along inversion lines are seen in association with CMEs, and form starting from flux cancellation at the photosphere (so-called bald patch reconnection, see e.g., \citealt{Green2011A&A...526A...2G}).
    As discussed in \citet{Aulanier2010ApJ...708..314A, Aulanier2014IAUS..300..184A}, tether-cutting coronal reconnection is also one of the key processes that builds flux ropes and brings them to critical heights where ideal instabilities (e.g. the torus instability) can be triggered. 
    This initial stage in the formation and/or destabilization of a magnetic flux rope produces the early $v_{nt}$ enhancement we detect, particularly in M-class events.
    \item \textbf{Gradual flux rope rise \& CME activation:} The early reconnection feeds magnetic flux into the rope and/or reduces overlying tension, causing a slow rise, as was shown for example in multiple observations of the so-called pre-eruptive hot magnetic flux rope by \citet{Cheng2020}. In this extended period of `precursor onset phase' in M-class flares, $v_{nt}$ continues to rise prior to the main flare phase, which could indicate a continuous destabilization of the magnetic configuration. A possible explanation is the continuous reconnection at the hyperbolic flux tube \citep{Demoulin1996JGR...101.7631D, Aulanier2005A&A...444..961A}, the 3D region where reconnection is the most likely to occur in the presence of a developed flux rope, 
    causing heating and slow rise of a flux rope~\citep{Woods2018ApJ...860..163W, Cheng2023ApJ...954L..47C, Xing2024ApJ...966...70X}. 
    \item \textbf{Instability threshold \& main flare onset:} The flux rope reaches a critical height, triggering runaway destabilization and acceleration (e.g., torus instability in \citealt{Zuccarello2015ApJ...814..126Z}; emergence flux destabilizing a flux rope in \citealt{Bamba2017ApJ...840..116B}) and the main flare energy release. This aligns with previous IRIS observations, then our primary $v_{nt}$ onset, and finally the GOES SXR onset across all flare classes. This progression, from the chromosphere and lower corona to higher temperatures, is consistent with our observations and has also been observed in other studies~\citep{Song2020ApJ...893L..13S,Panos2023A&A...671A..73P}.
\end{enumerate}

Hinode/EIS offers unique advantages over GOES integrated studies by spectroscopically resolving coronal plasma ($\sim$0.5--2.5 MK) through \ion{Fe}{8}--\ion{Fe}{16} lines, complementing the chromospheric and transition region studies by IRIS, and hot onset studies that primarily diagnose plasma above 10 MK. Our detection of consistent $v_{nt}$ enhancements 4 to 25 minutes before the SXR emission becomes detectable (Figure \ref{fig:NTV_onset_plot}, \ref{fig:NTV_onset_table}) highlights the fundamental role of non-thermal broadening in the primary energy release mechanisms.

\section{Conclusions}

In this paper, we analyze the flare footpoint non-thermal velocity of 1,449 flares observed by EIS from 2011 Jan 1 to 2024 Nov 9, representing the largest ever flare study using EIS spectral data. Our key findings include: 
\begin{enumerate}
    \item Non-thermal velocity consistently increases prior to the GOES SXR start time across all flare classes, peaks at the GOES SXR flare peak, and then gradually decreases during the decay phase.
    \item The primary non-thermal velocity enhancement precedes the GOES SXR start time by 4--25 minutes. Onset timing patterns vary with flare magnitude: smaller flares (C1, C2--3) show tendencies for temperature-dependent progression, while larger flares exhibit more compressed, near-simultaneous onsets. When focusing on well-sampled lines, onset times are remarkably consistent across all flare classes.
    \item The onset timing patterns suggest different energy release characteristics: smaller flares tend toward more sequential, gradual processes. However, these sequential timing patterns are likely within our sampling and analysis error, warranting further investigation with more precise instruments; while larger flares (C>4, M-class) exhibit more explosive energy deposition across multiple temperature regimes, consistent with the standard flare model.
    \item M-class flares exhibit an early `precursor onset' of non-thermal velocity enhancement $\sim$25--45 minutes before the SXR start time ($\sim$30--60 minutes before peak time), which is not consistently observed in C-class events.
    \item \review{CME-associated M-class flares show earlier and more uniform precursor onsets (45--74 minutes before peak) than non-CME flares, of which only some spectral lines display a precursor onset (31--54 minutes before peak). This suggests a strong link between extended pre-flare non-thermal broadening and successful eruptions}.
    \item The creation of this comprehensive EIS flare catalog (2011-2024) demonstrates the value of long-term, large-scale spectroscopic surveys for identifying and validating consistent pre-flare signatures beyond individual case studies.
\end{enumerate}

These findings establish an important connection between early non-thermal motions, subsequent heating, and ultimate energy release, providing important new constraints for understanding flare initiation and development. \review{The systematic detection of a precursor onset phase specifically in M-class flares suggests a signature linked to eruptive processes~\citep{Harra2013ApJ...774..122H,Chitta2025arXiv250312235C}, though we cannot rule out that similar mechanisms occur below our detection threshold in smaller events. This interpretation is consistent with recent modeling efforts demonstrating how magnetic avalanche processes produce the observed spectroscopic signatures~\citep[e.g.,][]{Cozzo2024A&A...689A.184C,Cozzo2025A&A...695A..40C}.}

A limitation of our study is the inherent bias toward \review{events with longer rise time}, which are more likely to be observed by EIS during their pre-flare phase. This selection effect means that our results cannot definitively determine whether pre-flare non-thermal velocity enhancements are equally common in \review{short rise-time flares}. Next-generation solar spectrometers could be important for testing these interpretations. MUSE will provide unprecedented temporal and spatial coverage of active regions in multiple temperature regimes, including very hot \ion{Fe}{19} -- \ion{Fe}{21} lines, while Solar-C EUVST will offer high-resolution spectroscopy from chromosphere to flare corona, allowing us to determine whether the early non-thermal broadening occurs simultaneously across different temperature regimes. This highlights the importance of extended missions that provide rich, long-term datasets for comprehensive statistical studies

This systematic, large-scaled observational understanding could be important for developing reliable flare prediction capabilities, and particularly for determining whether a flare will produce a CME. By establishing clear links between early non-thermal motions, subsequent heating, and ultimate energy release, our results provide both important constraints for flare models and a framework for future investigations with these next-generation instruments.

\section{Data availability}
The flare class-averaged non-thermal velocity curves and individual flare $v_{nt}$ values at each time step used in this study are publicly available on Zenodo at \href{https://doi.org/10.5281/zenodo.15613861}{doi:10.5281/zenodo.15613861}.

\acknowledgements{

ASHT and HE acknowledge support through the European Space Agency (ESA) Research Fellowship Programme in Space Science. ASHT and AB thank the Leiden-ESA LEAPS program, the result of the student project contributed to this work. ASHT also thank the ESA faculty research grant that supported the collaboration visit that cultivated this study. DB is funded under Solar Orbiter EUI Operations grant number ST/X002012/1 and Hinode Ops Continuation 2022-25 grant number ST/X002063/1. LAH is supported through a Royal Society-Research Ireland University Research Fellowship. PT is funded for this work by NASA contract NNM07AB07C (Hinode/XRT) to the Smithsonian Astrophysical Observatory, contract 8100002705 (IRIS), and  contract 4105785828 (MUSE) to the Smithsonian Astrophysical Observatory.  
JMS is supported for this work by contracts NNG09FA40C (IRIS) and 80GSFC21C0011 (MUSE). The work of DHB was performed under contract to the Naval Research Laboratory and was funded by the NASA Hinode program. ASHT thanks MSSL computing group and private conversation with Paul Prior. Hinode is a Japanese mission developed and launched by ISAS/JAXA, collaborating with NAOJ as a domestic partner, and
NASA and STFC (UK) as international partners. Scientific operation of Hinode is performed by the Hinode science team organized at ISAS/JAXA. This team mainly consists of scientists from institutes in the partner countries. Support for the post-launch operation is provided by JAXA and NAOJ (Japan), STFC (UK), NASA, ESA, and NSC (Norway).

This paper made use of several other open source packages including astropy~\citep{AstropyCollaboration2022ApJ...935..167A}, sunpy~\citep{Mumford2020JOSS....5.1832M, SunPyCommunity2020ApJ...890...68S}, matplotlib~\citep{Hunter2007CSE.....9...90H}, numpy~\citep{Harris2020Natur.585..357H}, scipy~\citep{Virtanen2020NatMe..17..261V}, EISPAC~\citep{Weberg2023JOSS....8.4914W}}.

\appendix
\section{GOES background removal example}\label{appendix:GOES_bacground_removal}

\begin{figure}

    \centering
    \includegraphics[width=\linewidth]{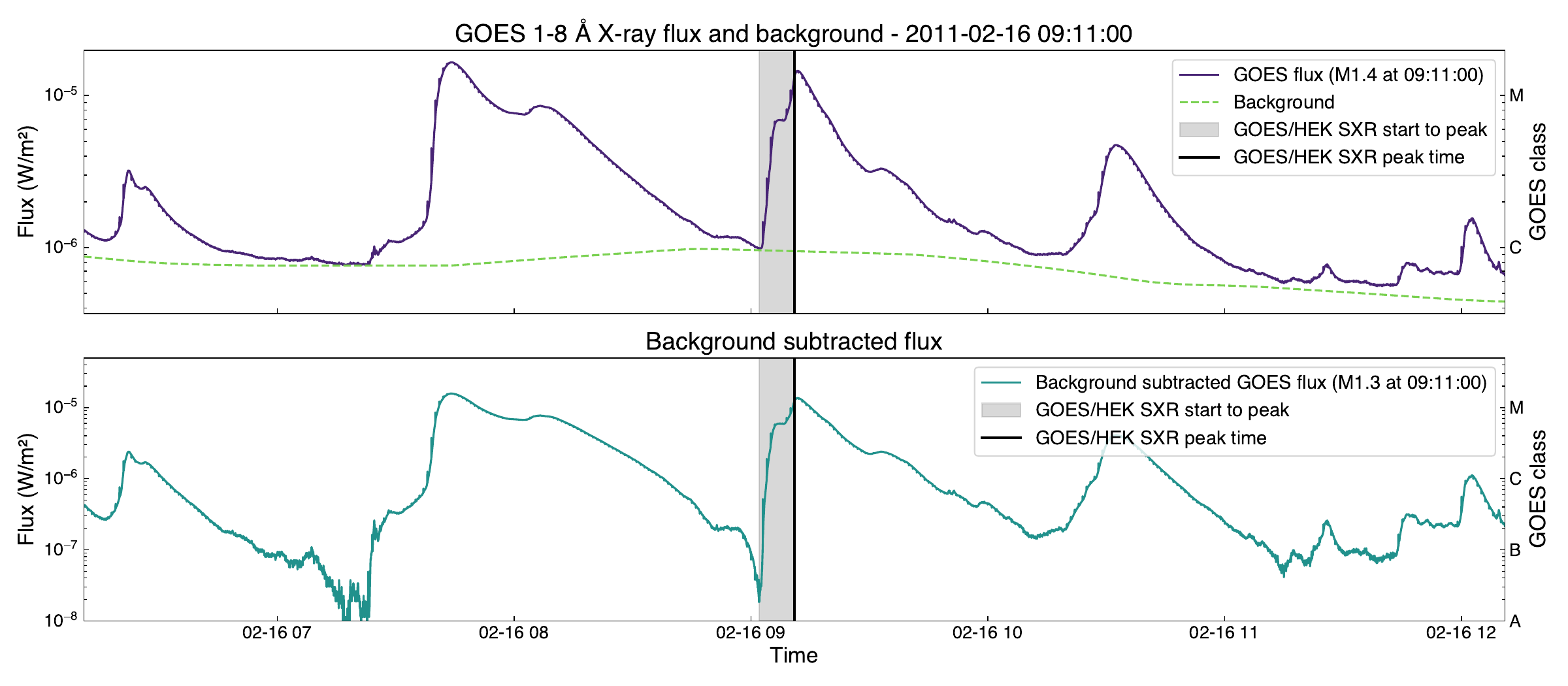}
    \caption{Example of GOES background removal algorithm illustrated in Section~\ref{sec:GOES_background_removal}. The purple line indicates the original GOES flux, light green indicates the rolling background, dark green at the bottom row indicates the background removed GOES flux. This figure shows a example of M1.4 flare being reclassified to M1.3 after background removal.}
    \label{fig:GOES_processing}
\end{figure}

Figure~\ref{fig:GOES_processing} shows an example of the GOES background removal used in reclassifying the GOES class.



\section{Non-thermal velocity results for C1, C2--3, C>4, and X-class flares}\label{appendix:multipanel}
Similar to Figure~\ref{fig:M_class}, this section shows the $v_{nt}$ evolution for C1, C2--3, C>4 and X-class flares. X-class flares show less spectral lines due to the limited amount of observations.

\begin{figure*}

    \centering
    \includegraphics[width=\linewidth]{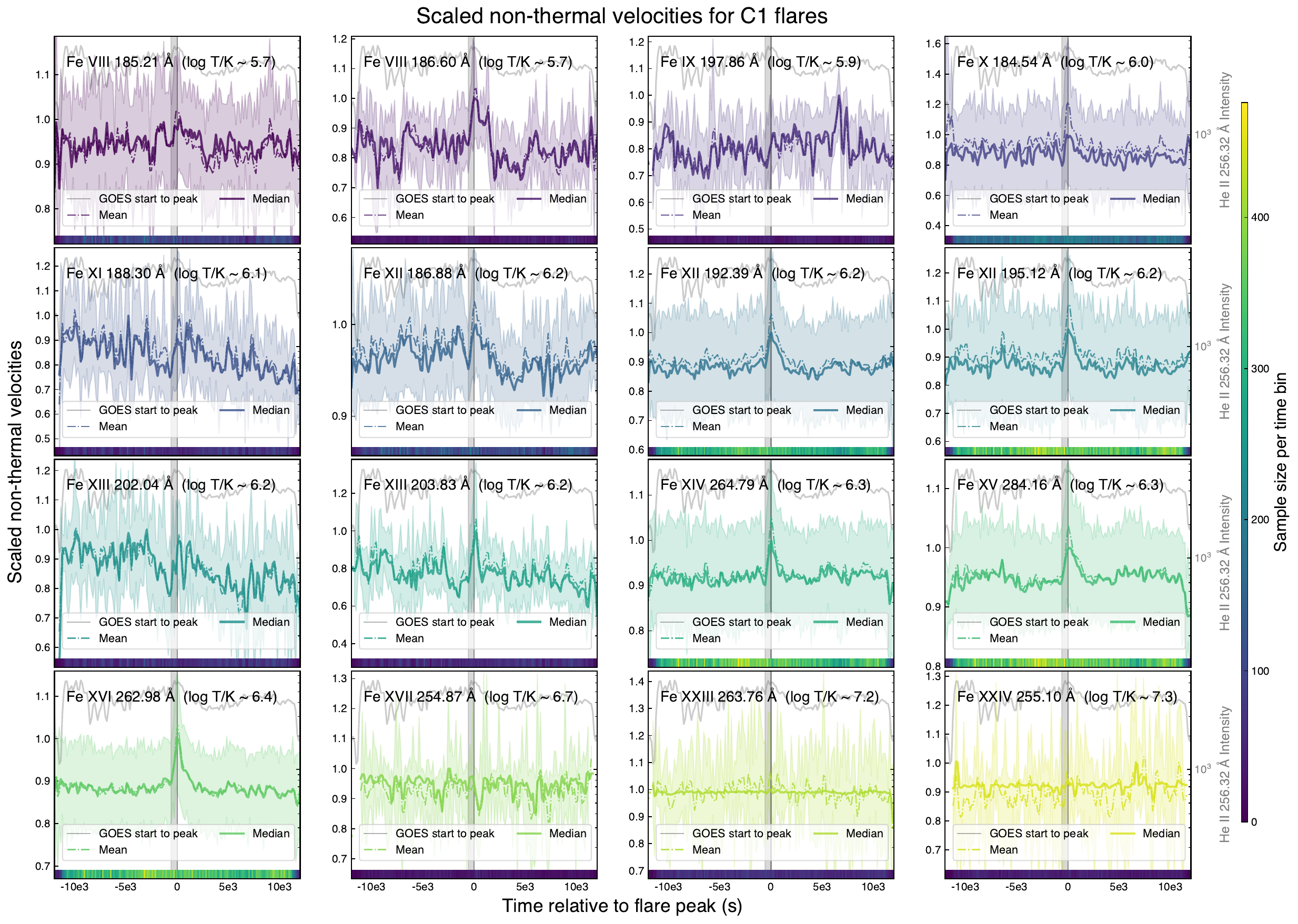}
    \caption{Similar plot to Figure~\ref{fig:M_class}. Non-thermal velocity results for C1 flares. Solid line indicates the median, dotted line inidcate the mean, and the shaded area indicates the 25 and 75 percentile. The vertical shaded area indicates the SXR start to peak time for this flare category. The horizontal bar at the bottom of each subplot indicate how vertical EIS columns went into the calculation of the respective bin.}
    \label{fig:C1}
\end{figure*}

\begin{figure*}
    \centering
    \includegraphics[width=\linewidth]{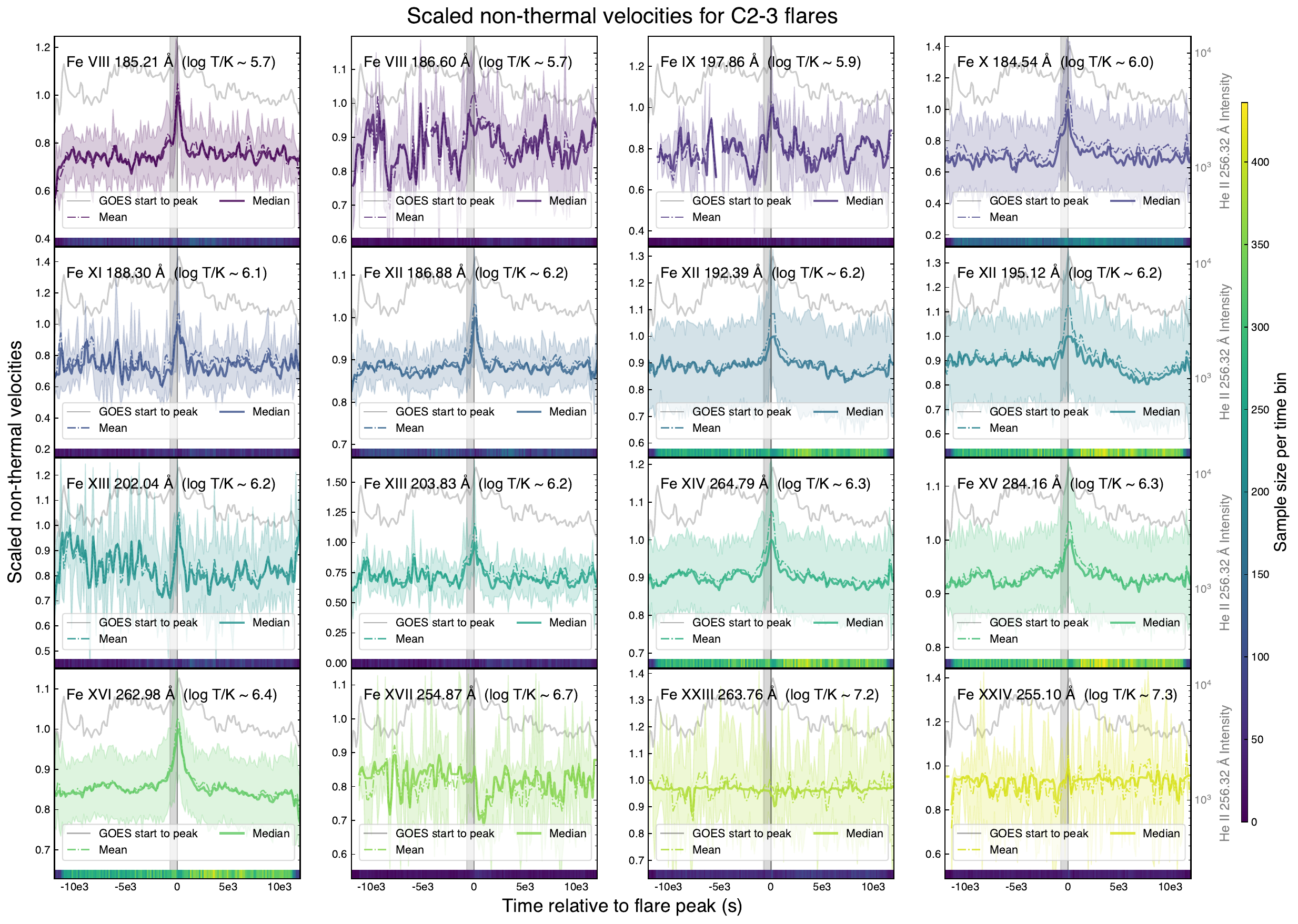}
    \caption{Similar to Figure~\ref{fig:C1} but on C2-3 flares.}
    \label{fig:C_below_C5}
\end{figure*}

\begin{figure*}

    \centering
    \includegraphics[width=\linewidth]{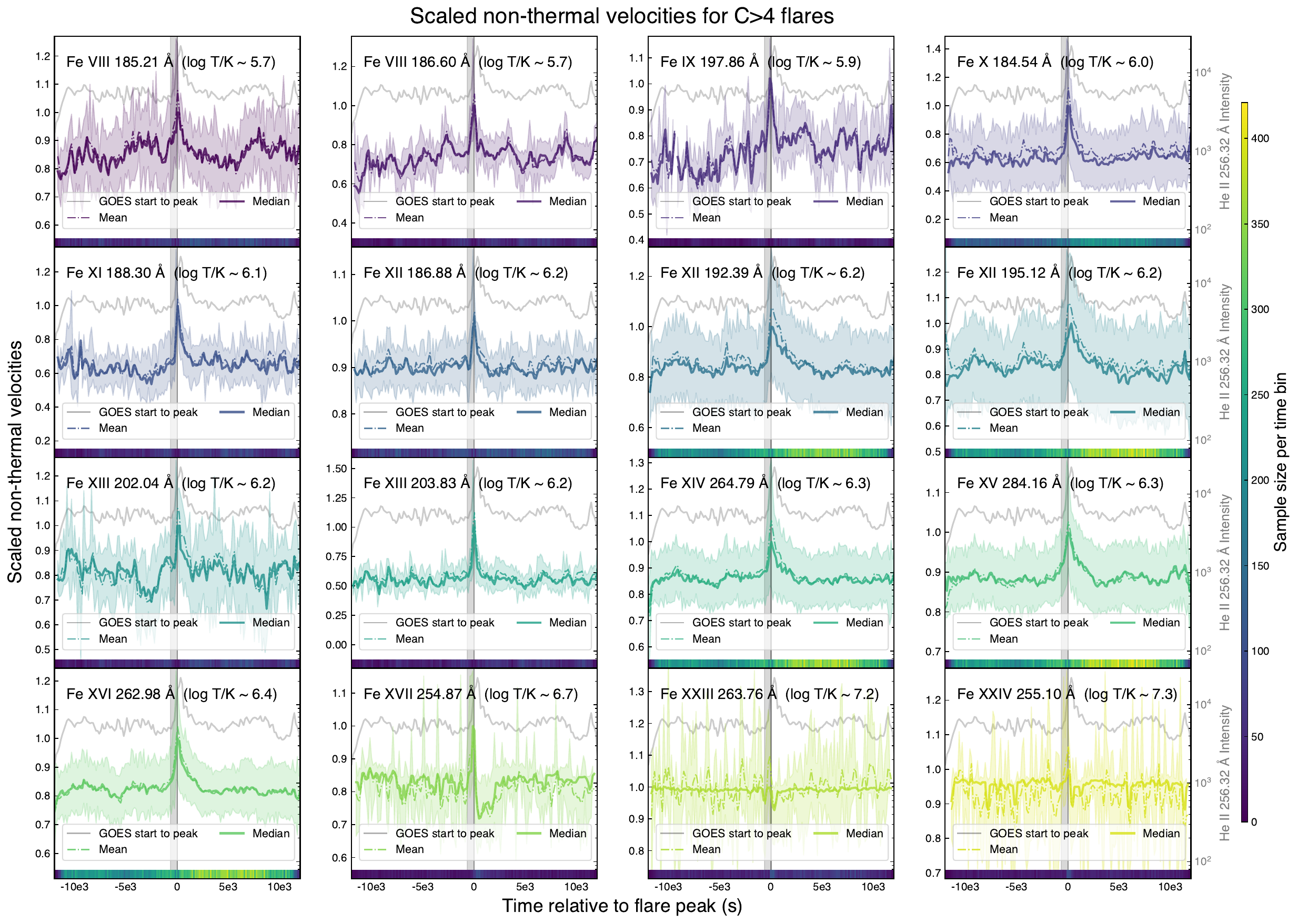}
    \caption{Similar to Figure~\ref{fig:C1} but on C$>4$ class flares.}
    \label{fig:C_above_C5}
\end{figure*}
\begin{figure*}

    \centering
    \includegraphics[width=\linewidth]{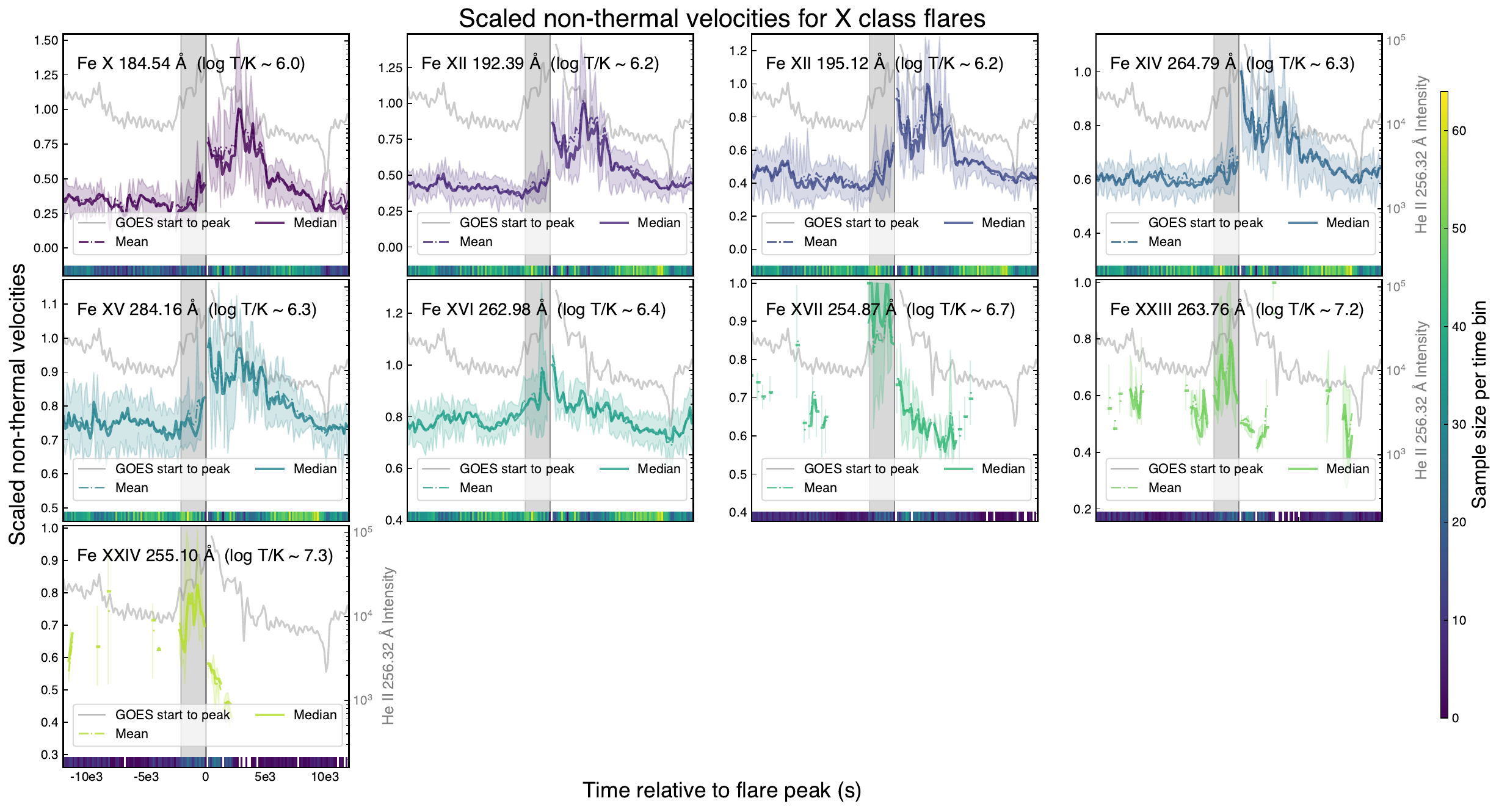}
    \caption{Similar to Figure~\ref{fig:C1} but on X-class flares.}
    \label{fig:X_class}
\end{figure*}

\bibliography{bib}{}
\bibliographystyle{aasjournal}



\end{document}